\documentclass[a4paper,fleqn]{cas-dc}
\usepackage[numbers]{natbib}
\usepackage{caption}
\graphicspath{{./images/}}
\begin{document}
\let\WriteBookmarks\relax
\def\floatpagepagefraction{1}
\def\textpagefraction{.001}

\shorttitle{Evaluation of Live Forensic Techniques in Ransomware Attack Mitigation}
\shortauthors{S.R.Davies,R.Macfarlane,W.J.Buchanan}

\title [mode = title]{Evaluation of Live Forensic Techniques in Ransomware Attack Mitigation}                      
\author[1]{Simon R. Davies}[type=editor,
                        auid=000,bioid=1,
                        ]
\cormark[1]
\ead{s.davies@napier.ac.uk}
\address[1]{School of Computing, Edinburgh Napier University, Edinburgh, UK}

\author[1]{Richard Macfarlane}[%
   suffix=,
   ]

\author[1]{William J. Buchanan}[%
   orcid=0000-0003-0809-3523]


\begin{abstract}
Ransomware continues to grow in both scale, cost, complexity and impact since its initial discovery nearly 30 years ago. Security practitioners are engaged in a continual "arms race" with the ransomware developers attempting to defend their digital infrastructure against such attacks. Recent manifestations of ransomware have started to employ a hybrid combination of symmetric and asymmetric encryption to encode user's files. 

This paper describes an investigation that tried to determine if the techniques currently employed in the field of digital forensics could be leveraged to discover the encryption keys used by these types of malicious software thus mitigating the effects of a ransomware attack. 

Memory was captured from a system infected by ransomware and its contents was examined using live forensic tools, with the intent of identifying the symmetric encryption keys being used. NotPetya, Bad Rabbit and Phobos hybrid ransomware samples were tested during the investigation. If keys were discovered, the following two steps were also performed. Firstly, a timeline was manually created by combining data from multiple sources to illustrate the ransomware's behaviour as well as showing when the encryption keys were present in memory and how long they remained there. Secondly, an attempt was made to decrypt the files encrypted by the ransomware using the found keys. In all cases, the investigation was able to confirm that it was possible to identify the encryption keys used. A description of how these found keys were then used to successfully decrypt files that had been encrypted during the execution of the ransomware is also given.

The resulting generated timelines provided a excellent way to visualise the behaviour of the ransomware and the encryption key management practices it employed, and from a forensic investigation and possible mitigation point of view, when the encryption keys are in memory.

\end{abstract}

\begin{keywords}
Live Forensics \sep Ransomware
\sep
\sep
\sep
----------------------------------------------
\sep
Article history
\sep
Received 1\textsuperscript{st} February 2020
\sep
Accepted 5\textsuperscript{th} April 2020
\sep
Available
\end{keywords}

\maketitle
\section{Introduction}
\label{cha:intro}
The aim of this paper was to investigate if the techniques commonly used in live forensics can be applied to the analysis of ransomware and in so doing allow the investigator to discover useful cryptographic fragments from the malware that could later be used to possibly reverse the effects of the ransomware's execution. While ransomware first appeared more than 30 years ago \cite{Salvi2015}, its initial impact on the computing community was small with only a few people being affected and recovery from the attack being trivial. 

The threat landscape changed in 2013 with the release of the CryptoLocker ransomware \cite{Bradley2016} where attackers adopted the three new technologies of cryptocurrency, TOR onion routing and cryptography. Combining them to produce a new breed of ransomware programs that have become more effective and aggressive than anything previously experienced. These new sophisticated attacks have generated large amounts of money for the perpetrators. Culminating in two of the biggest ransomware attacks in recent times, WannaCry which is estimated to have cost \$8 billion and NotPetya which is estimated to have cost \$10 billion \cite{Mekynyk2019}. The number of malware attacks is generally considered to be growing year-on-year \cite{Europol2016,Intelligence2019}. 

There exists a separate research field concerning itself with the study and development of live forensics techniques and more specifically with live memory analysis. Some research performed in this field has been into the recovery of cryptographic fragments, specifically encryption keys, from the contents of the system's memory where cryptographic processes are active. A lot of this research has proven very successful \cite{Balogh2011,Maartmann-Moe2009} allowing the researchers to determine the encryption keys used by the cryptographic programs and subsequently using them to decrypt files. However, no specific research has been found where these techniques have been applied to machines that have ransomware active on them.

The main contributions of this research is first to provide confirmation that forensic techniques are able to be employed to investigate the volatile memory of machines infected by ransomware and can be leveraged to extract useful cryptographic fragments that can be used to mitigate the effects of the attack. Secondly time lining the behaviour of the ransomware attack and highlighting when and for how long the encryption keys are available for extraction.

The rest of the paper is structured as follows. Section 2 - A review and discussions of related work in these areas. Section 3 - Description of the design philosophy. Section 4 - Description of the experiment implementation and the results achieved. Section 5 - Critical analysis of the experimental results and comparison to similar work in the field. Section 6 –Discussion of the findings.

\section{Related Work}

Ransomware is one of the most widespread and damaging threats that internet users face today \cite{Sophos2019} and is often classified by the type of encryption used \cite{Al-rimy2018}:

\begin{itemize}
\item \textbf{Symmetric Crypto-Ransomware (SCR)} uses one key for both encryption and decryption allowing the attack to complete in a shorter time, reducing the chances of it being discovered.
\item \textbf{Asymmetric Crypto-Ransomware (ACR)} uses different keys for encryption and decryption. 
\item \textbf{Hybrid Key Crypto-Ransomware (HCR)} firstly uses symmetric encryption to encrypt the user’s files as fast as possible. After which the symmetric key is encrypted using asymmetric encryption.
\end{itemize}
Historically the incidents of ransomware have been increasing year on year prompting Interpol to declare in 2016 that ransomware had become \emph{the most prominent malware threat [\ldots] for citizens and enterprises alike} \cite{Europol2016} and MalwareBytes \cite{Malwarebytes2019} reporting a 500\% year on year increase in attacks demonstrating that the trend upwards is set to continue. Europol confirming in 2018 that they believe that ransomware will retain its dominance for several years to come \cite{Europol2018}. 

Even at the time of writing American government agencies are struggling with the effects of a recent ransomware attack \cite{ODonnall2019} and US authorities are preparing for similar attacks during the voter registration for the 2020 elections \cite{Hautala2019}.

 The NotPetya cyber attack is considered the costliest attack in history \cite{Kapersky2018} with an estimated cost of \$10 billion, whereas WannaCry, according to various estimates, lies in the \$4–\$8 billion range.

\subsection{Live forensics and memory acquisition}
Static forensic analysis methods are used in analysing evidence from a computer system that has been turned off. The problem with this approach is that significant information stored in the computers volatile memory is lost when the machine is switched off. Examples of information that could be present in memory are encryption keys, open connection details, running processes, logged-in users, and so on \cite{Bashir2013}. To address this issue a complimentary forensics approach known as live forensics has been developed. Live forensic analysis primarily targets the computers volatile data which can only be acquired from a running system. When applied to ransomware analysis, the live forensics techniques can be complimented by combining them with malware analysis techniques. 

One important aspect of live forensics is the examination of the systems memory where the malicious code is running. This examination is normally performed off-line so that the contents of the memory are not affected by the examination or the memory capture tools used. To achieve this, the memory of the system to be examined needs to be captured and saved. 

According to Ruff\cite{Ruff2008} there are three main memory capturing techniques:
\begin{enumerate}
    \item Software-based, which typically involves executing extraction programs. An issue with this approach being that the execution of software would impact the contents of the captured systems memory.
    \item Hardware-based, which typically involves connecting devices, such as PCMCIA cards or USB sticks and are not always practical in live scenarios as physical access to the machine is required \cite{McLaren2019}.
    \item Virtualization technology-based techniques.
    \end{enumerate}{}

A detailed compilation of the techniques available is provided in \cite{Carvey2009} along with advantages and disadvantages of each approach. 

Using the virtualization approach, a snapshot of the analysed system’s volatile memory is extracted using tools provided by the virtualization software. This snapshot is then inspected by an analyst using a variety of specialised forensic tools\cite{Nissim2019}. Obviously to use this technique the system must be running in a virtualised environment. The advantages of this approach being that no trace of any extraction program exists in the captured memory and any running malicious programs are unaware that they are being analysed or that the memory dump was taken  \cite{Dinaburg2008,Ligh2014, McLaren2019a}.

The challenge of memory acquisition in this context is to discover cryptographic artefacts, such as the encryption keys, in a manner that allows the target device to continue to operate normally, while the memory is being acquired \cite{McLaren2019}.

\subsection{Cryptanalysis live forensics}
Some work has been previously performed into the possibility of using live forensic techniques to discover encryption keys that may be present in a computers memory. It was not possible to find any literature that focused specifically on ransomware in particular, however similar work on key determination in volatile memory has been performed for SSH tunnels, encrypted volumes, WinRAR, WinZip and Skype. \cite{Balogh2011,Maartmann-Moe2009,McLaren2019}.

Several research papers confirm the assumption that for a system to be able to encrypt/decrypt data, then the cryptographic algorithm needs to have access the encryption keys and these are normally held in volatile memory. Balogh \cite{Balogh2011} state that encryption in real-time is only performed in memory which means that the encryption keys must also be present there. So in the case of symmetric encryption it means that the keys also needed for decryption will also be recoverable from memory. With regards to key management, Maartmann-Moe \cite{Maartmann-Moe2009} state that it is clear that cryptographic keys need to be present in memory during encryption when using standard computer hardware. 

In extensive tests conducted on 10 different cryptographic systems the researchers \cite{Maartmann-Moe2009} were always able to retrieve all the cryptographic keys from memory for every application tested using their specifically developed tool called \emph{interrogate}. While these researchers have not investigated ransomware specifically, their findings strongly suggest that it would be possible to extract ransomware cryptographic keys using similar techniques.

\subsubsection{Examination of memory methods}
The usual process for locating something is to try to identify some characteristic of what is being located and then to look for that characteristic. One characteristic of cryptographic keys is that they are usually chosen at random. Most code and data is not chosen at random and it turns out that this differentiation is significant \cite{Shamir1998}. When data is random it has higher entropy than patterned information. This means that it should be possible to locate cryptographic keys among other data by locating sections with unusually high entropy \cite{Balogh2011}. The authors found that the memory block where the main and the auxiliary AES keys are located has a recognizable structure and high entropy.

In reality, symmetric cryptographic keys are just short sequences of random looking data, often 16–32 bytes long residing amongst other pieces of data with a much lower entropy. 

\subsubsection{Identifying keys in memory}
In order to extract encryption keys from memory, they must first be identified \cite{Halderman2009}. Several researchers have discovered \cite{Halderman2009,Ptacek2008,Shamir1998}, that encryption keys in memory are far more structured than previously believed and several strategies to locate the keys have been proposed and are discussed below.

Using a high-entropy searching approach as suggested by Shamir \cite{Shamir1998} and tested by \cite{Maartmann-Moe2009}. Since it is known that key data has a higher entropy than non-key data, one way to locate a key is to divide the data into small sections, measure the entropy of each section and display the locations where there is particularly high entropy \cite{Shamir1998}. 
 
Search for certain known patterns in the memory such as key schedule which are specific to certain types of encryption \cite{Halderman2009,Ptacek2008}. These patterns could also consist of known memory offsets, specific lengths of high entropy memory locations, or known patterns of entropy.

\subsubsection{Identifying AES keys}
From a review of current ransomware samples it has been determined that the majority of modern crypto ransomware are now hybrid in nature (HCR) \cite{Al-rimy2018}.
The public key of the asymmetric encryption being delivered with the ransomware, while the private key is retained by the attacker. As the private key of the asymmetric encryption is never present on the machine, this paper will concentrate on the identification of the key used during the symmetric encryption phase of the ransomware’s execution which in the majority of cases is the Advanced Encryption Standard (AES) key.

AES is a Substitution-Permutation (SP)-network based cipher that works on 128-bit blocks, and can use either 128, 198 or 256 bit keys. It is considered by some researchers to be virtually unbreakable \cite{Saravanan2014} and impossible to decrypt without the correct key \cite{Maartmann-Moe2009}. 

\begin{figure}[]
\centering
\includegraphics[width=3.5in]{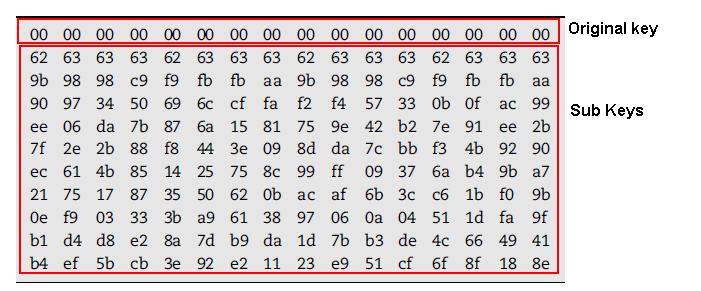}
\caption{AES Key and Key Schedule \protect\cite{Maartmann-Moe2009}.}
\label{fig:AESSchedule}
\end{figure}

Modern symmetric key cryptosystems are constructed by repeatedly applying a simpler function where several iterations or “rounds,” are done. From the master key, a derivation function, derives different sub-keys used in each round. This is known as the key schedule algorithm and has been stated by many researchers that this information needs to be present in memory \cite{Balogh2011,Hargreaves2008,Maartmann-Moe2009}. This knowledge of the cryptosystem can be used to search for this key pattern within the memory \cite{Balogh2011}. The search criteria being that there is a mathematical relationship between the master key and sub-keys. The key schedule is often computed ahead of time, in what appears to be a security-performance trade-off, and kept in the memory while encryption/decryption is performed \cite{Maartmann-Moe2009}. An example of an 128-bit empty AES key (all zeros)  and its associated key schedule \cite{Maartmann-Moe2009}, extracted from memory is shown in Fig.~\ref{fig:AESSchedule}.

Notably, the key schedule for a 128 bit AES key is represented as a flat array of bytes in memory, where the first 16 bytes (or 128 bits) constitutes the original key. The remaining 160 bytes are the round keys derived from this key. For larger AES keys, the corresponding key schedule is also larger.

Several tools have been developed that use this phenomenon to identify AES keys in memory such as AESFinder \cite{Halderman2009, Heninger2008}, Volatools \cite{Walters2007}, interrogate \cite{Maartmann-Moe2009} and Findaes \cite{Kornblum}, however no research has been found to suggest that they have ever been used to analyse cryptographic ransomware in particular. 

\subsubsection{Method}
Some of the key works in the area have used the same common experimental method   
\cite{Balogh2011,Hargreaves2008,Maartmann-Moe2009,Nissim2019,Walters2007}. Firstly running the program under investigation in a virtual environment, then using tools from the virtual environment to capture memory dumps of the systems memory in a secured and trusted manner. Once the required number of memory captures has been completed they are then analysed.

Several researchers including  \cite{Balogh2011,Halderman2009,Maartmann-Moe2009,McLaren2019a,McLaren2019} have had success in extracting the encryption keys  through the discovery of cryptographic information in volatile memory.

One thing to remember when applying this method to the ransomware samples analysed in this report is that these samples perform several steps before they actually begin encrypting data on the victims system \cite{Nissim2019}. So unlike the programs investigated by other researchers the ransomware encryption keys will not be immediately present in the memory when the programs starts executing and determining when to capture the memory becomes critical to the success of the experiments. Ideally the memory should be captured while the ransomware is encrypting files. The encryption process can last from several minutes to a few hours, and the program may perform good key management on the completion of encryption by removing the keys from memory, so determining the point when the memory capture should be performed is crucial. The technique described uses volatile memory analysis combined with empirical observations rather than focusing on specific API calls that the ransomware may  employ.

\section{Design}
\label{cha:design}
\subsection{Environment design}
In order to conduct valid, realistic ransomware experiments the test victim's machine needs to mimic a real machine as much as possible with any private or confidential information removed\cite{Hoopes2009} and ideally isolated from the internet \cite{Ahmad2016}.
Based on the research performed by Bose \cite{Bose2018}, it was decided to implement the test environment for this project using the VirtualBox virtualization software provided by Oracle.

When designing a test environment, one of the key recommendations from Rossow \cite{Rossow2012} was that it should be as realistic as possible. Using these guidelines as well as the three points raised by Sanabria \cite{Sanabria2007}, the test environment was designed to contain three virtual machines. 
Two of these virtual machines were victim test machines. Only one of which were ever active at any given time and it is on these where the ransomware was executed. 

In all but the most basic experiments at least one other system is required to provide network support services \cite{Sanabria2007} for the victim, as denying all-network access to the sample under analysis will most likely result in incomplete observations of the malware’s behaviour \cite{Egele2012}. Therefore a common technique \cite{Sikorski2012} was used where a third virtual machine was present on the virtual network to provide network services to the victim machines such as DNS, IRC, HTTP as well as handling possible requests made by the malware back to its command and control (C\&C) server \cite{Sanabria2007}.

These machines were connected via a ‘host-only’ virtual network connection, and they were the only machines on this virtual network. This configuration provides complete isolation of these guest machines from the host machine and thus the host’s network connections. The physical host machine consisted of a laptop which itself was “air gapped,” thus providing a second layer of isolation. Having both network access and content on the victim virtual machine contributed significantly to it appearing to the malware as a real machine, encouraging the malware to behave normally. Appropriate containment policies such as firewall and anti-virus were also deployed on the host machine \cite{Rossow2012}.

\subsection{Experiment design}
At its most abstract level the designed experiments could be considered as follows. A ransomware sample is executed within a virtual environment. During this execution, copies of the machines volatile memory are taken. Forensic tools are then used to analyse these captured memory files in an attempt to discover the encryption key. The found keys are then used to decrypt files encrypted during the ransomware's execution to confirm that the correct symmetric encryption key has been identified. 



This investigation can be broken down in to three separate sub experiments. The results of which, when combined were used to validate or disprove the hypothesis that live forensic techniques could be used to mitigate a ransomware attack. Each round of experiments began with a fresh version of a virtual machine that reflected a realistic workstation being started and an example of the ransomware under investigation being executed. The following three experiments were  then performed.

\subsubsection{Experiment part 1 - Identify the key in memory}
\begin{figure}[]
\centering
\includegraphics[width=2.5in]{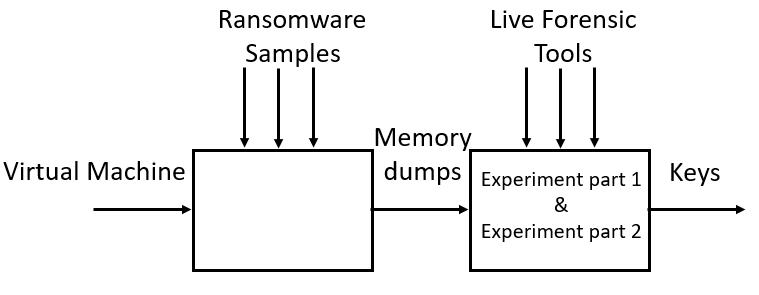}
\caption{Overview of Experiment part 1 \& Experiment part 2.}
\label{fig:metaexperiment12}
\end{figure}

 A memory dump from the machine where the ransomware was being executed was captured. It was known during selection of the ransomware samples for these experiments that AES was being used for the symmetric part of the encryption, so once the dump had been completed it was analysed using live forensics tools to determine if the AES key can be discovered. Special attention being paid to areas of memory that exhibited high entropy. A graphic representation of this and the following experiment is shown in Fig.~\ref{fig:metaexperiment12}.

\subsubsection{Experiment part 2 - Key time line creation}
A memory dump was taken at regular intervals during the complete execution life cycle of the ransomware to determine at what point the key is loaded into memory and for how long it remains there. This aids the execution of experiment part 1 by determining when to execute point A in Fig.~\ref{fig:method16}. An outcome of this experiment was an approximate timeline for the execution of the ransomware.

\subsubsection{Experiment part 3 - Validate found keys}
\begin{figure}[]
\centering
\includegraphics[width=2.5in]{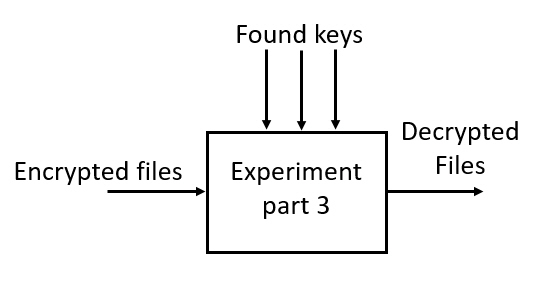}
\caption{Overview of Experiment part 3.}
\label{fig:metaexperiment13}
\end{figure}

If any keys were discovered during experiment part 1, they were tested to determine if they could decrypt any of the control files encrypted by the ransomware. As indicated by points B and C in Fig.~\ref{fig:method16}. A tool developed by the author using the Python programming language was used to perform the decryption attempt. A graphic representation of this experiment is shown in Fig.~\ref{fig:metaexperiment13}.

\subsubsection{Combined experiment}
\begin{figure}[]
\centering
\includegraphics[width=2.5in]{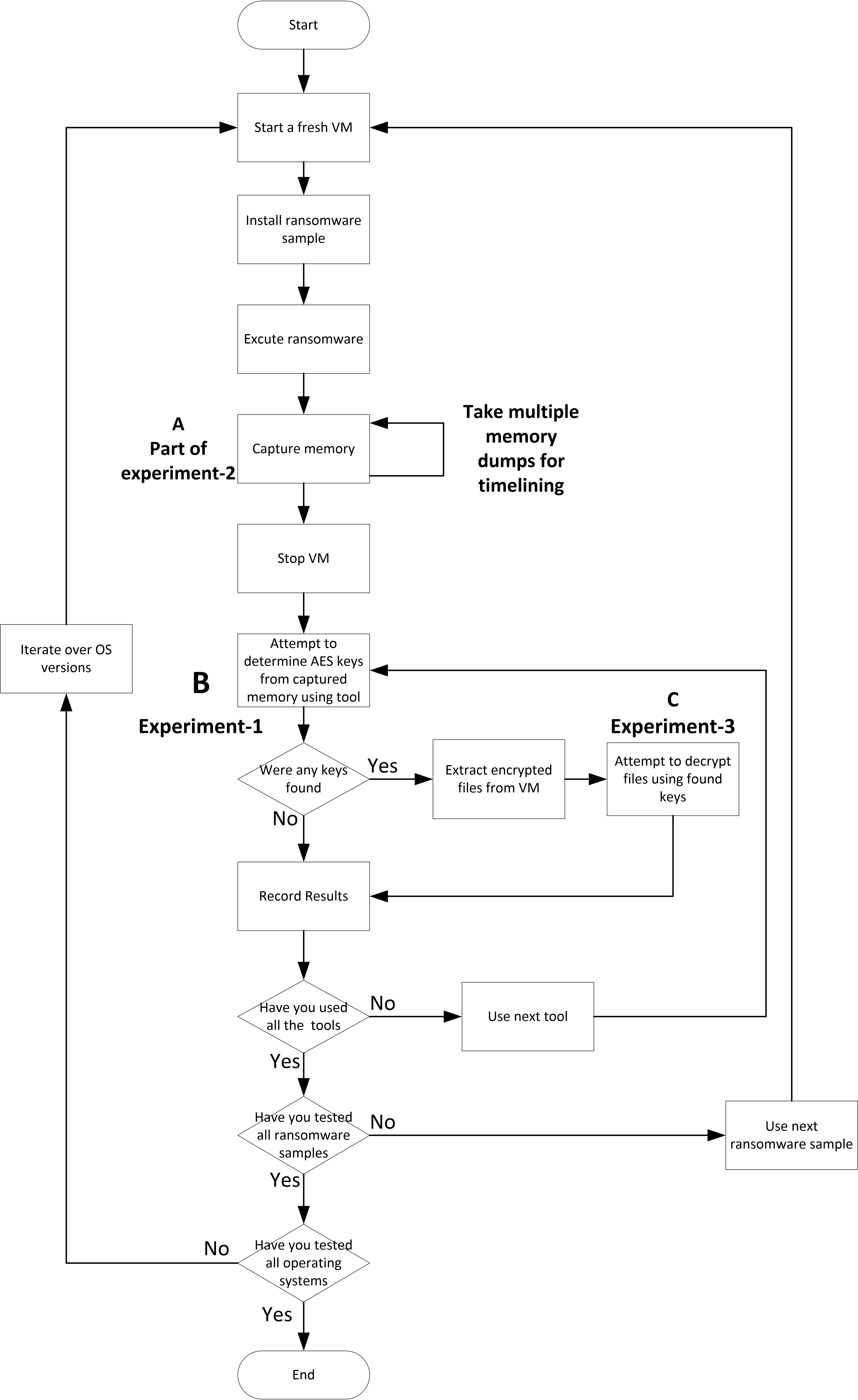}
\caption{Experiment flow.}
\label{fig:method16}
\end{figure}

The overall experimental suite was based on similar methods identified during the literature review \cite{Balogh2011,Hargreaves2008,Maartmann-Moe2009,Nissim2019,Walters2007}.
While the designed experiments in this investigation have significant similarities with previous work such as using windows operating systems on virtual machines and using similar tools for key extraction. These experiments differ in that multiple key extraction tools on multiple operating systems were tested and the discovered keys were checked to confirm that they successfully decrypted the control files. Also multiple memory dumps were taken during the experiment to allow for the creation of a timeline for the ransomware's execution. A graphical representation of the overall experiment is given in Fig.~\ref{fig:method16} and the main steps are discussed below.

\noindent\textit{Iterate over Operating Systems versions.}
When discussing realism in experimentation Rossow \cite{Rossow2012} cautioned against performing experiments on just one operating system and then drawing general conclusions. To guard against this potential criticism, the experiments performed in this research were conducted against the top two most commonly used versions of the windows operating systems currently in use.

\noindent\textit{Start a fresh VM.}
Results are only comparable if each sample is executed in an identical environment \cite{Hoopes2009} so a fresh VM was started at the beginning of each experiment. 

\noindent\textit{Install Ransomware.}
The sample to be tested was extracted from the archive and prepared for execution.

\noindent\textit{Execute Ransomware.}
The chosen ransomware sample was executed from the command line.

\noindent\textit{Capture Memory.}
The timings of when to take a copy of the working memory of the virtual machine was determined by the outcome of experiment part 2. If the keys became available in memory, then a copy of the machines memory was taken, using tools provided by the virtualization software in similar techniques used by other researchers \cite{McLaren2019a,Nissim2019}.

\noindent\textit{Stop VM.}
Once the required number of memory captures has been performed, or if the ransomware had completed, then the victim virtual machine is halted.

\noindent\textit{Attempt to determine AES keys from captured memory.}
Analysis was performed on the captured memory samples with the aim of identifying candidate AES keys, using the three tools discussed below:
\begin{enumerate}
    \item \textbf{findaes} - A tool developed by Kornblum \cite{Kornblum} based on the work by Trenholme \cite{Halderman2009, Trenholme} and tries to find the keys using the AES key schedule structure \cite{Trenholme2014}.
    \item \textbf{interrogate} - A tool developed by Maartmann-Moe \cite{Maartmann-Moe2009} also based on the work by Trendholme\cite{Trenholme}.
    \item \textbf{RansomAES} - A hybrid tool developed by the author which incorporates logic from the Volatility Framework \cite{Volatility2019} together with the logic from findaes in an attempt to improve the accuracy and performance of the key detection.
\end{enumerate}{}

\noindent\textit{Extract Encrypted Files.}
A set of typically targeted  \cite{CERT-EU2017} control files with known content were placed on the victim machine prior to the ransomware's execution. Once the ransomware had executed, these files were analysed to determine if they had been encrypted. 

\noindent\textit{Attempt to decrypt files.}
If any candidate keys were discovered during the analysis stage of the experiment, they were then used to try and decrypt the encrypted control files extracted from the victim virtual machine. The AES Initialisation Vector (IV) required for decryption, were contained within the encrypted file and used in combination with the candidate keys to decrypt the file.

\section{Implementation and results}
\label{cha:implementation}
\subsection{Ransomware sample selection}
Recent ransomware attacks were researched \cite{Comodo2018, Hautala2019, Kapersky2018, ODonnall2019,Vanderburg2019}
and based on the findings,  it was decided to select the following three recent ransomware samples for analysis. All of which used AES for the symmetric portion of the encryption. 

\noindent\textbf{NotPetya}. This ransomware attack is considered to be the most damaging attack ever \cite{Kapersky2018}.

\noindent\textbf{Bad Rabbit}. An adaptation of the NotPetya ransomware family that emerged in 2018 \cite{SonicWall2019}. 

\noindent\textbf{Phobos}. One of the most recent ransomware samples found where detailed information is available \cite{Issa} and also one of the most prevalent in Q4 2019 \cite{ODonnell2020}. 

Specific details of the ransomware samples used were validated by VirusTotal (www.virustotal.com) are described in Table~\ref{tab:ransomwaresamples}. The three chosen ransomware samples were similar to each other in that they all used AES symmetric encryption and also all used the same key for all files encrypted

\begin{table}
\caption {Ransomware Samples} \label{tab:ransomwaresamples} 
\begin{tabular}{ l l}
\toprule
Name & SHA256 Checksum\\
\midrule
NotPetya & \scriptsize\makecell[l]{027cc450ef5f8c5f653329641ec1fed91f694e0d22992896\\3b30f6b0d7d3a745} \\
Bad Rabbit & \scriptsize\makecell[l]{630325cac09ac3fab908f903e3b00d0dadd5fdaa0875ed8\\496fcbb97a558d0da} \\
Phobos & \scriptsize\makecell[l]{a91491f45b851a07f91ba5a200967921bf796d38677786\\de51a4a8fe5ddeafd2} \\
\bottomrule
\end{tabular}

\end{table}

\subsubsection{Other ransomware}
The following ransomware samples were initially considered before being rejected.

\begin{itemize}
\item Wannacry. Testing confirmed what was detailed in the literature that this strain used unique AES keys for each encrypted file\cite{Berry2017,VipreSecurity2017}. After conducting multiple tests of the memory acquired during the execution of this malware using all the live forensics tools, no recoverable AES keys were found.
\item Cerber. This ransomware appears not to use AES encryption.
\item Lucky/nmare. The sample of this ransomware required access to the internet to be able to download the file encryption modules as they are not delivered with the initial sample. The services provided by the Debian virtual machine were not able to trick the ransomware into executing normally and it was deemed too risky to allow this external network access.
\item Satan, SamSam and GrandCab. It was not possible to trigger these samples of ransomware to encrypt any of the control the files or display the ransom message.
\end{itemize}

\subsection{Test machine configuration}
The laptop used for testing had no external network connections and as an added precaution was set to airplane mode and had the wifi card switched off. When discussing the virtual environments this physical machine is referred to as the host machine as it hosts the virtual environment within VirtualBox.

Three guest machines were defined in the virtual environment running on the laptop. Two victim machines with different operating systems on each, used to test the behaviour of the ransomware code and one network services machine that was used to provide any network services. Details of the virtual guest machines are given in Table ~\ref{tab:virtualconfig}. 

\begin{table}
\caption{Virtual Hardware Configurations}
\begin{tabular}{ l l l}
\toprule
\makecell[l]{Machine Name} & Operating System & Purpose\\
\midrule
Windows 7 & \makecell[l]{Windows 7 Ultimate \\ Build 7600} & \makecell[l]{Ransomware \\Victim  Machine} \\ \\
Windows 10 & \makecell[l]{Windows 10 Pro \\ Build 10586.494} & \makecell[l]{Ransomware \\Victim  Machine} \\ \\
Debian & Debian 5.2.9 &  Network Services\\
\bottomrule
\end{tabular}

\label{tab:virtualconfig}
\end{table}

The guest machines were connected together using the recommended  ‘host-only’ connection technique 
\cite{Hoopes2009},creating a separate virtual LAN providing isolation and containment of the guest machines. The virtual LAN and the guest machines were run in two separate configurations depending on which windows version was being tested.

The following tools were used during the configuration of the Debian machine. \textit{fakedns.py –} used to provide DNS services to the network and \textit{INetSim -} Considered to be the best free tool \cite{Sikorski2012} for providing fake network service emulation creating the illusion of a realistic pseudo network that the malware can interact with if required.

\subsection{Experiments}
Details of the experiments performed are given below
\subsubsection{Experiment part 1 - Identify the key in memory}
A fresh VM was started and a copy of the machines memory was taken prior to the execution of the ransomware, so that any AES keys that are present in the machines memory prior to the execution of the ransomware could be identified and excluded from the experiments results. To aid experiment reproducibility, the commands used to launch each of the ransomware samples is provided below:

\noindent\textit{For NotPetya}
\begin{verbatim}
   c:\windows\system32\rundll32.exe 
     c:\ransomware\netpetya.dll, #1 30
\end{verbatim}

\noindent\textit{For BadRabbit}
\begin{verbatim}
   c:\ransomware\sample.badrabbit1.bin.exe
\end{verbatim}

\noindent\textit{For Phobos}
\begin{verbatim}
   c:\ransomware\1saas.bin.exe
\end{verbatim}

After waiting for a short period, a copy of the guest's machines memory was taken. The length of waiting time varied between ten seconds and two minutes depending on the ransomware strain. The time required to wait was determined through trial and error. To capture the memory, the following command was executed on the host machine:

\begin{verbatim}
VBoxManage.exe debugvm <VBox Machine Name> 
      dumpvmcore --filename <filename>.elf
\end{verbatim}

The memory dump file was then analysed by each of the selected live forensics memory tools. Again to aid reproducibility, the commands used are given below:
\begin{verbatim}
findaes <filename>.elf
interrogate -a aes -k 128 <filename>.elf
ransomaes  -p <ransomware pid> -t Win7SP0x86 <filename>.elf
\end{verbatim}

Any keys found resulting from the execution of these tools were recorded and used as input for experiment part 3. If no keys were found then the experiment was extended in one minute intervals and new memory dumps taken. The experiment terminated when keys were found or the execution of the ransomware completed.

\subsubsection{Experiment part 2 - Key time line creation}
If keys were discovered in experiment part 1, then this experiment was also performed. Memory dumps are taken regularly throughout the execution time frame of the ransomware. The time interval used between memory dumps varied depending on the ransomware sample and was determined via trial and error over multiple executions. 
The dumps were analysed using one of the selected tools to confirm that they keys were still present. The times when the keys were present was recorded and a basic timeline for the ransomware execution was created. The timeline creation process was predominantly a manual task, combining the results recorded from the experiments, observations of system behaviour, descriptions gained from the literature review and utilisation of the six step ransomware model \cite{McAfeeLabs2016}.

\subsubsection{Experiment part 3 - Validate found keys}
If candidate AES keys were discovered in experiment part 1, then these were used in an attempt to decrypt the control files. This was partially an automated task with some manual steps. A tool \verb|decrypt.py| was created by the author to perform the basic AES decryption using the \verb|'AES'| cipher object from the \verb|'Crypto.Cipher'| python library.


The program was able to determine the required IV from the supplied encrypted file and then use this together with the discovered candidate keys to try and decrypt the file. Determination if the file was correctly decrypted remained a manual task. The resulting decrypted file normally required extra modifications such as adding a header or removing a trailer.

\section{Results and Discussion}
\label{cha:evaluation}
This chapter is divided into separate sections, one for each of the ransomware samples tested. Each of these sections discusses the results of the three experiments conducted.

\subsection{NotPetya} \label{sectionnp}
The execution of the ransomware appeared to follow the descriptions provided by \cite{Berry2017,VipreSecurity2017}. The main steps being:
\begin{enumerate}
    \item Adding persistence and gathering user's credentials.
    \item Scanning the machine for files to encrypt. This sample seems to ignore the control files with the 'txt' extension.
    \item Encrypting the identified files using AES encryption with what appears to be the same AES key being used for all the files. Interestingly neither the filename or the file meta information changes when the file is encrypted.
    \item Attempting lateral movement to other machines, however no actual evidence of this was discovered.
    \item After 1 hour, rebooting the machine automatically.
    \item Displaying a fake chkdsk command output, while encrypting the Mater Boot Record.
    \item Once the fake chkdsk completes, the machine reboots automatically again.
    \item After reboot, the ransomware message shown in Fig.~\ref{fig:npransom} is displayed.
        \begin{figure}[]
        \centering
        \includegraphics[width=3in]{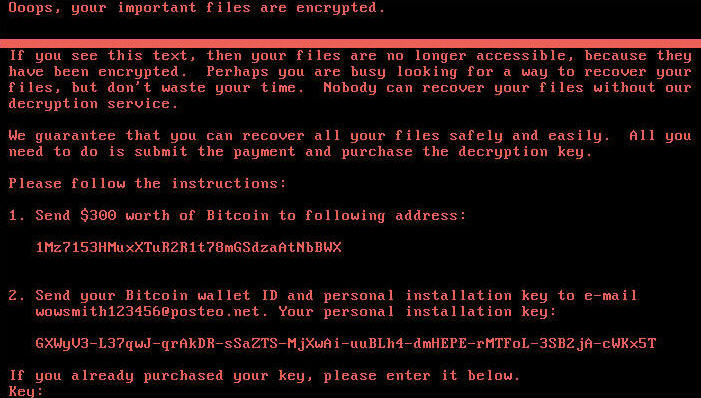}
        \caption{NotPetya Ransom Message.}
        \label{fig:npransom}
        \end{figure}
\end{enumerate}

\subsubsection{Experiment part 1 - Identify the key in memory}
All three live forensics tools used to examine the memory dumps were able to successfully identify AES keys in the memory of the ransomware process.
 
 \subsubsection{Experiment part 2 - Key time line creation}
A total of fifteen memory dumps were taken and then analysed to determine if they contained AES keys. It was identified that the key became available within 2 minutes of the start of the ransomware, and remained in memory until the machine was automatically rebooted by the ransomware after 60 minutes. The key did not survive the reboot and was not recoverable from memory after this. A graphical representation of some of the ransomware's behaviour and key availability is shown below in Fig~\ref{fig:nptimeline}.

\begin{figure}[]
\centering
\includegraphics[width=3.5in]{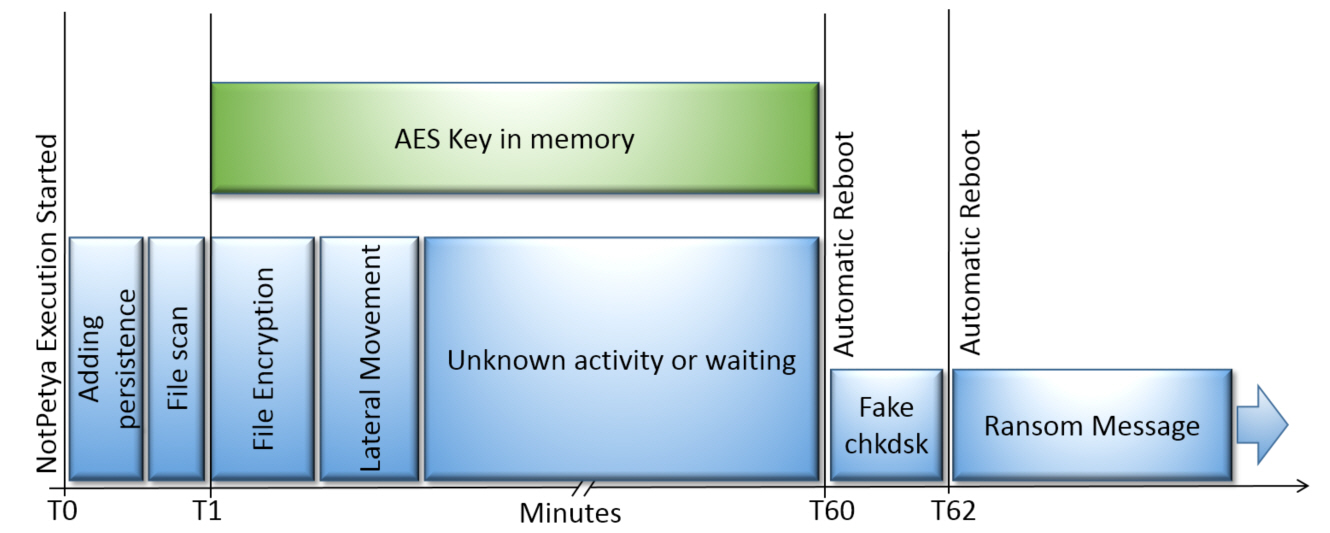}
\caption{NotPetya Timeline.}
\label{fig:nptimeline}
\end{figure}

This timeline correlates well with the findings of other researchers \cite{Berry2017,VipreSecurity2017}. However no research was found that analysed when and for how long the actual key remains in memory. Using this diagram it is clear to see that it is present for a total 59 minutes, which is the majority of the ransomware's execution time. Also no similar graphical representation of the ransomware time line was found in the literature, the one shown in Fig.~\ref{fig:nptimeline} being generated by the author.

\subsubsection{Experiment part 3 - Validate found keys}\label{notpetyadecrypt}
When the NotPetya ransomware encrypts a file, the first 16 bytes of the file are overwritten with the AES Initialisation Vector (IV) value \cite{Sood2017}. A program was developed that firstly reads the IV value from the encrypted file, then used this together with the key found in experiment part 1 to decrypt the files contents. 

Using this technique pdf, doc, docx, xls and xlsx extracted control files were successfully recovered using the same AES key. Each of these file types required different headers to be inserted and some files required that some bytes be removed from the end of the file. 

\subsection{Bad Rabbit}\label{sectionbr}
The execution of this ransomware followed the description provided by \cite{2017MalwarebytesLABS2017,Mamedov,Perekalin2018} and is similar to the steps used by the NotPetya ransomware. The main steps being:

\begin{enumerate}
    \item Adding persistence.
    \item Scanning the machine for files to encrypt. This sample seems to ignore control files with the 'txt' and 'jpg' file extensions.
    \item Encrypting the identified files using AES encryption.
    \item After 14 minutes a ransom note file is created on the C drive.
    \item One minute later the machine is automatically rebooted.
    \item The machine restarts with what appears to be a normal windows desktop. However in the background the MBR is being encrypted.
    \item 22 minutes after the initial execution of the ransomware, the machine automatically reboots again.
    \item After reboot, the ransomware message shown in Fig.~\ref{fig:brransom} is displayed and the user is prevented from accessing their windows installation.
        \begin{figure}[]
        \centering
        \includegraphics[width=3in]{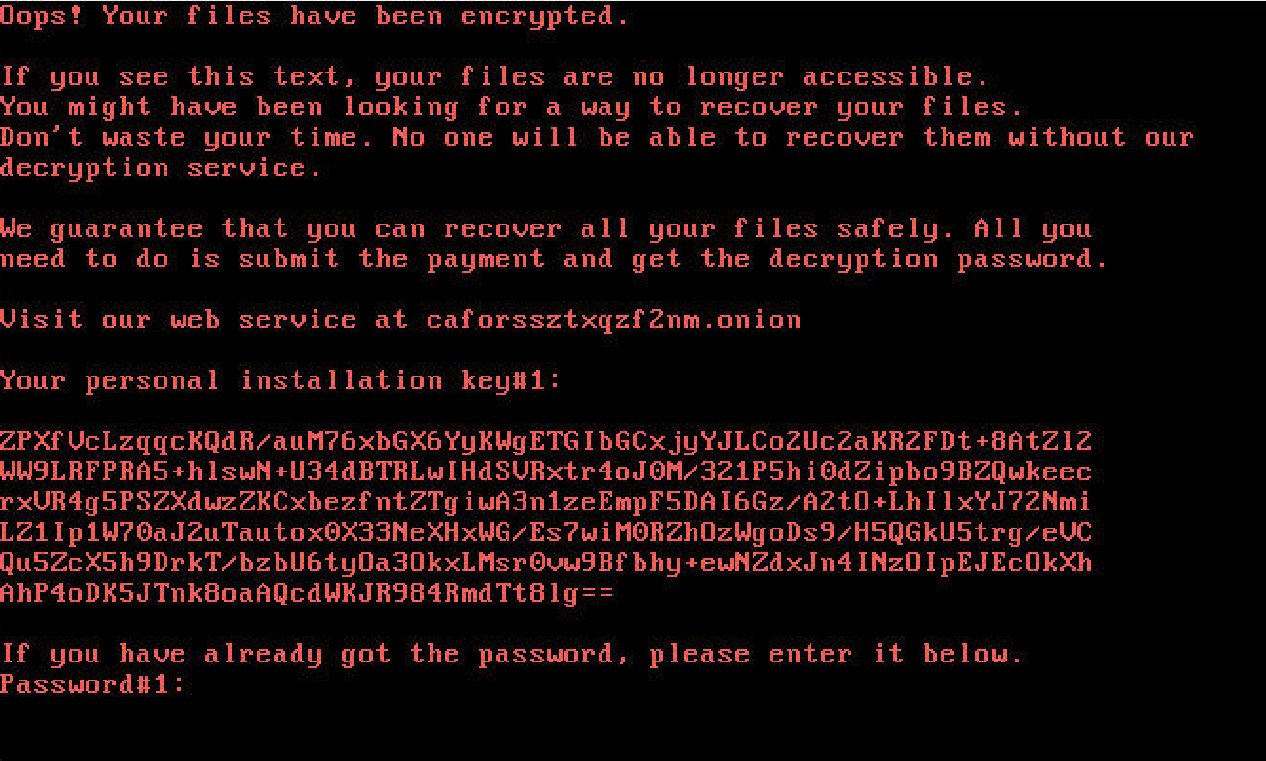}
        \caption{Bad Rabbit Ransom Note.}
        \label{fig:brransom}
        \end{figure}
\end{enumerate}

\subsubsection{Experiment part 1 - Identify the key in memory}
All three live forensics tools used to examine the memory dumps were able to identify AES keys in the memory of the ransomware process.
 
 \subsubsection{Experiment part 2 - Key time line creation}
Again a total of fifteen memory dumps were taken and analysed to determine if they contained AES keys. It was identified that the key became available within 1 minute of the start of the ransomware execution, and only remained in memory while the encryption was being done. An approximation of this being 30 seconds. The key did not appear again even after the machine reboot. A graphical representation of some of the ransomware's behaviour and key availability is shown below in Fig.~\ref{fig:brtimeline}.
\begin{figure}[]
\centering
\includegraphics[width=3.5in]{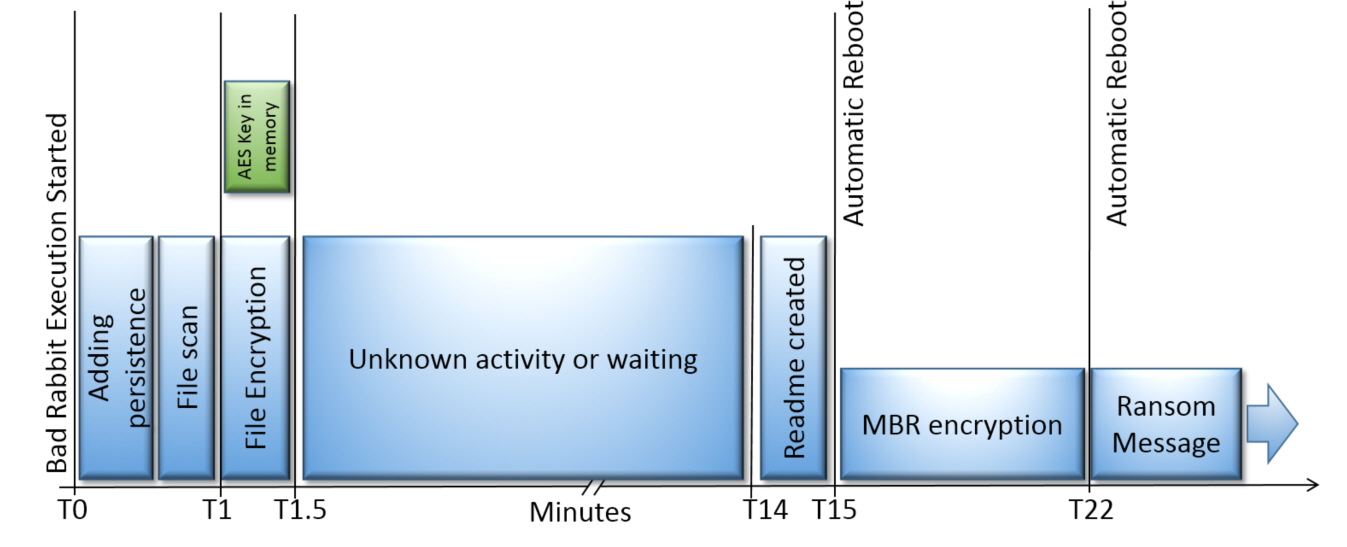}
\caption{Bad Rabbit Timeline.}
\label{fig:brtimeline}
\end{figure}
This timeline matches the description given by \cite{2017MalwarebytesLABS2017}. Using the time line diagram it is clear to see that it is only present for 30 seconds which is a fraction of the overall execution time and much shorter than the NotPetya ransomware. The key could possibly be present at other times, but missed due to the sampling period. No similar graphical representation of the ransomware time line was found in the literature, the one shown in Fig.~\ref{fig:brtimeline} being generated by the author.
 
\subsubsection{Experiment part 3 - Validate found keys}
Files are encrypted using the same format as the NotPetya ransomware and the steps described in section \ref{notpetyadecrypt} can be used to decrypt the files encrypted with the Bad Rabbit ransomware. 
Using this technique pdf, doc, docx, xls and xlsx files were successfully recovered using the same AES key. 

\subsection{Phobos}\label{sectionph}
The execution of the ransomware followed the description provided by \cite{Issa}. The main steps being:
\begin{enumerate}
    \item Adding persistence and gathering credentials.
    \item Scanning the machine for files to encrypt. This sample encrypted all the control files.
    \item Encrypting the identified files using AES encryption with what appears to be the same AES key being used for all the files initially encrypted. The file names were also changed.
    \item After approximately 2 minutes the ransom note shown in Fig.~\ref{fig:phransom} is displayed.
        \begin{figure}[]
        \centering
        \includegraphics[width=3.5in]{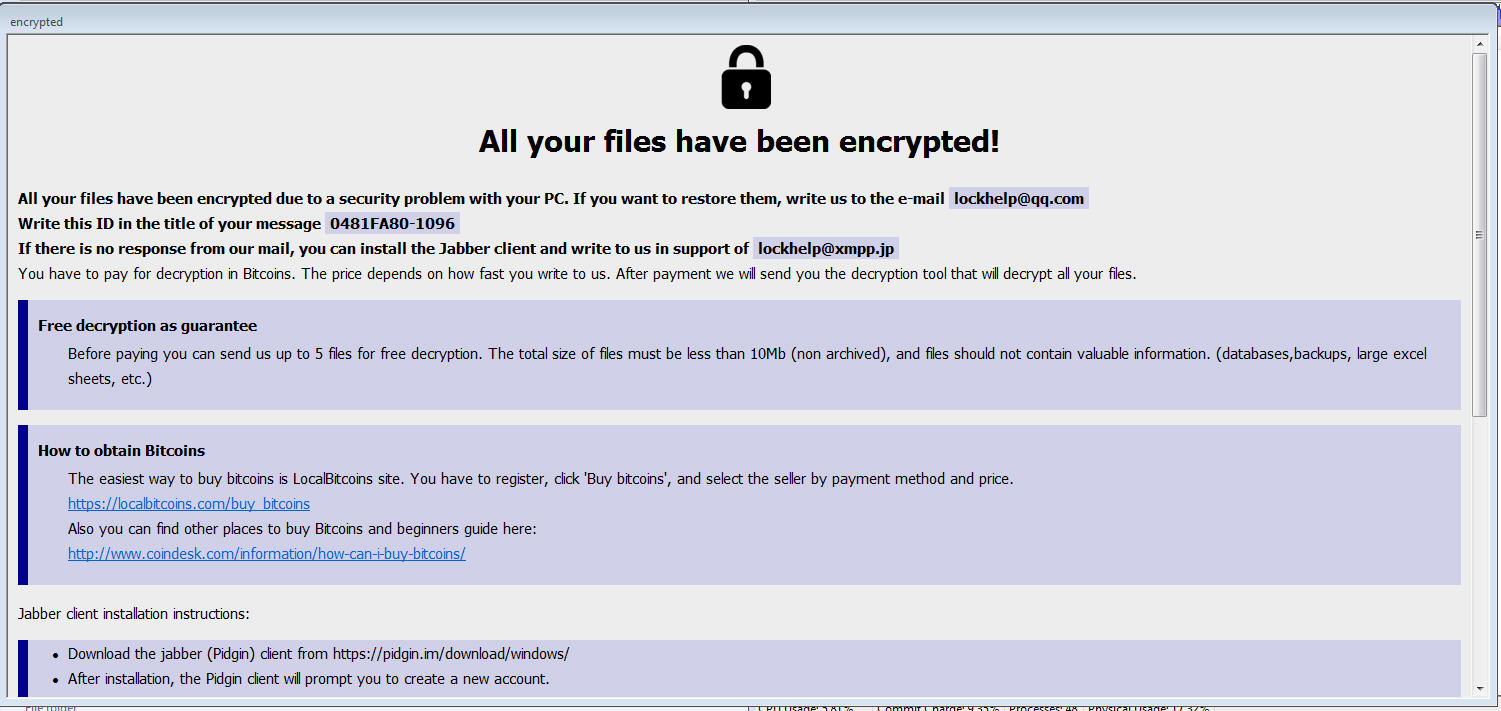}
        \caption{Phobos Ransom Note.}
        \label{fig:phransom}
        \end{figure}
    \item Interestingly the machine still operates to some extent and any new files created are encrypted using a different AES key. The researcher was not able to discover this secondary AES key. This could be an area of further research it is believed that these keys should also be present.
    \item The machine does not reboot automatically. If the user reboots the machine, then the same ransomware message is displayed. 
\end{enumerate}

\subsubsection{Experiment part 1 - Identify the key in memory}
All three live forensics tools used to examine the ransomware process memory were able to identify the 256 bit AES key used by the ransomware to encrypt the files. 
 
 \subsubsection{Experiment part 2 - Key time line creation}
\begin{figure}[]
\centering
\includegraphics[width=3.5in]{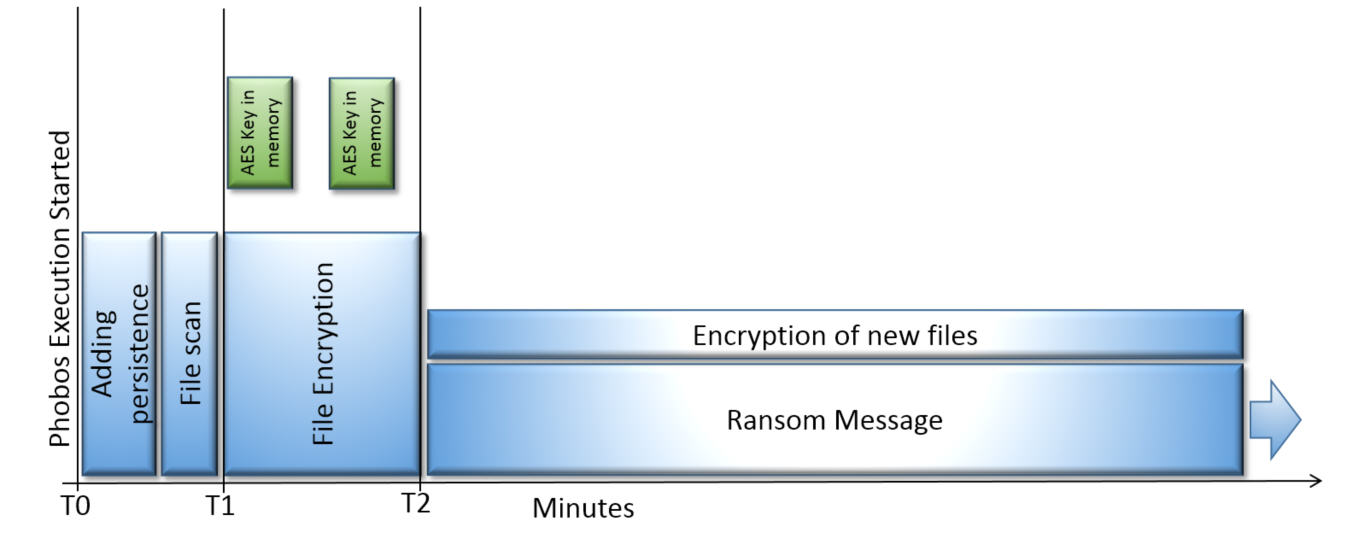}
\caption{Phobos Timeline.}
\label{fig:phtimeline}
\end{figure}

Similarly  a total of fifteen memory dumps were taken and  analysed to determine if they contained AES keys. It was identified that the key initially became available within 1 minute of the start of the ransomware execution. The same AES key was loaded in to memory and removed several times during this initial encryption of the machine. The key was erased when the ransom message was displayed. The ransomware continued to encrypt any new files created, using a different AES key. Several unsuccessful attempts were made to try and capture this secondary key from memory. As with the previous two ransomware samples no AES key survives a machine reboot. A graphical representation of some of the ransomware's behaviour and key availability is shown in Fig.~\ref{fig:phtimeline}.

This timeline correlates well with the description given by \cite{Issa, Security2017}. Using this diagram it can be easily seen that the same AES key is present in memory on several different occasions. It is also worth bearing in mind that there could be occasions where the keys presence was missed due to the sampling period. 
No similar graphical representation of the ransomware time line was found in the literature, the one shown in Fig.~\ref{fig:phtimeline} being generated by the author.

\subsubsection{Experiment part 3 - Validate found keys}
Additional information is added to the end of a file that has been encrypted by the Phobos ransomware. This extra information includes some padding, followed by what is believed to be the AES IV value, then followed by 128 bytes that is the same for all the encrypted files. A detailed description of the encrypted file appears in the work by Issa \cite{Issa} who hypothesis that this 128 byte block could be the encrypted asymmetric key. There is also a fixed string at the end of the block, in this case it is 'LOCK96' but other versions of Phobos have been observed with different keywords such as 'DAT260'.

A program was developed that firstly reads the IV value from the encrypted file, then uses this together with the key found in experiment part 1 to decrypt the file. Using this technique pdf, doc, docx, xls and xlsx files were successfully recovered using the same AES key.

\section{Conclusion}
\label{cha:conclusion}
Building on other work in the field of live forensics  \cite{Balogh2011,Hargreaves2008,Maartmann-Moe2009}, AES key recovery from volatile memory techniques were successfully used to identify encryption keys being used by ransomware samples. As research in this specific area was not identified, techniques and tools used for TrueCrypt and Skype encryption key recovery were used to identify keys for recent, high impact ransomware samples. Secondly the work aimed to attempt to evaluate whether consistent timelines of when these keys where stored in memory could be generated. 


Similar execution times and results were recorded for the findaes and RansomAES live forensic tools, The RansomAES tool created by the author had similar results as the pure findaes tool, indicating that the added extra functionality for ransomware did not result in an improvement in performance. The time taken for the interrogate tool to complete was almost 100 times longer. Apart from this extended execution time all three tools were shown to successfully identify AES keys being used by three ransomware samples chosen for the experiments on both windows 7 and windows 10 operating systems. This supports the hypotheses that live forensic techniques could be used to mitigate a ransomware attack.
 The authors believe that the increased execution time of the interrogate tool could be caused by the implementation of the logic this tool used to identify the AES key as fundamentally it is based on the same research as findaes \cite{Trenholme2014,Trenholme}. Interrogate seems not to take into account the entropy of the candidate keys prior to calculation of the schedule, rather it calculates the schedule for all candidate keys irrespective their entropy and this may be one reason for the impact in performance.

However these results support the hypothesis with some caveats from a forensic investigation point of view. These being that the machine has not been rebooted since the commencement of the attack, and also that the memory capture is performed near the start of the ransomware's execution, to ensure that the required keys are still present in memory. Also this technique cannot be applied to all ransomware families, as demonstrated by the analysis of the Wannacry ransomware sample, where  AES keys were not able to be identified. 

It was also observed that in case of a ransomware attack where the user was still able to interact with the operating system after the initial encryption has completed, such as the Phobos ransomware, a second AES key looks to be in memory. This key is being used to perform any subsequent encryption, such as on any new files created. This approach being possibly utilised to frustrate the recovery of the original key and protect the original encryption. The keys found during the work seem to align with suggestions from previous work \cite{Balogh2011, Hargreaves2008}, which indicate that it was likely that these secondary AES keys in theory should also be available in memory. Further research is planned to analyse various ransomware samples to determine if capture of these secondary keys is also possible and useful in a forensic investigation.

\printcredits

\bibliographystyle{cas-model2-names}

\bibliography{bibliography}

\begin{thebibliography}{54}
\expandafter\ifx\csname natexlab\endcsname\relax\def\natexlab#1{#1}\fi
\providecommand{\url}[1]{\texttt{#1}}
\providecommand{\href}[2]{#2}
\providecommand{\path}[1]{#1}
\providecommand{\DOIprefix}{doi:}
\providecommand{\ArXivprefix}{arXiv:}
\providecommand{\URLprefix}{URL: }
\providecommand{\Pubmedprefix}{pmid:}
\providecommand{\doi}[1]{\href{http://dx.doi.org/#1}{\path{#1}}}
\providecommand{\Pubmed}[1]{\href{pmid:#1}{\path{#1}}}
\providecommand{\bibinfo}[2]{#2}
\ifx\xfnm\relax \def\xfnm[#1]{\unskip,\space#1}\fi
\bibitem[{{2017 Malwarebytes LABS}(2017)}]{2017MalwarebytesLABS2017}
\bibinfo{author}{{2017 Malwarebytes LABS}}, \bibinfo{year}{2017}.
\newblock \bibinfo{title}{{Bad Rabbit: a closer look at the new version of
  Petya/NotPetya}}.
\newblock \URLprefix
  \url{https://blog.malwarebytes.com/threat-analysis/2017/10/badrabbit-closer-look-new-version-petyanotpetya/}.
\bibitem[{Ahmad et~al.(2016)Ahmad, Woodhead and Gan}]{Ahmad2016}
\bibinfo{author}{Ahmad, M.A.}, \bibinfo{author}{Woodhead, S.},
  \bibinfo{author}{Gan, D.}, \bibinfo{year}{2016}.
\newblock \bibinfo{title}{{The V-network testbed for malware analysis}}.
\newblock \bibinfo{journal}{Proceedings of 2016 International Conference on
  Advanced Communication Control and Computing Technologies, ICACCCT 2016} ,
  \bibinfo{pages}{629--635}\DOIprefix\doi{10.1109/ICACCCT.2016.7831716}.
\bibitem[{Al-rimy et~al.(2018)Al-rimy, Maarof and Shaid}]{Al-rimy2018}
\bibinfo{author}{Al-rimy, B.A.S.}, \bibinfo{author}{Maarof, M.A.},
  \bibinfo{author}{Shaid, S.Z.M.}, \bibinfo{year}{2018}.
\newblock \bibinfo{title}{{Ransomware threat success factors, taxonomy, and
  countermeasures: A survey and research directions}}.
\newblock \bibinfo{journal}{Computers and Security} \bibinfo{volume}{74},
  \bibinfo{pages}{144--166}.
\newblock \URLprefix \url{https://doi.org/10.1016/j.cose.2018.01.001},
  \DOIprefix\doi{10.1016/j.cose.2018.01.001}.
\bibitem[{Balogh and Pondelik(2011)}]{Balogh2011}
\bibinfo{author}{Balogh, {\v{S}}.}, \bibinfo{author}{Pondelik, M.},
  \bibinfo{year}{2011}.
\newblock \bibinfo{title}{{Capturing encryption keys for digital analysis}}.
\newblock \bibinfo{journal}{Proceedings of the 6th IEEE International
  Conference on Intelligent Data Acquisition and Advanced Computing Systems:
  Technology and Applications, IDAACS'2011} \bibinfo{volume}{2},
  \bibinfo{pages}{759--763}.
\newblock \DOIprefix\doi{10.1109/IDAACS.2011.6072872}.
\bibitem[{Bashir and Khan(2013)}]{Bashir2013}
\bibinfo{author}{Bashir, M.S.}, \bibinfo{author}{Khan, M.N.A.},
  \bibinfo{year}{2013}.
\newblock \bibinfo{title}{{Triage in Live Digital Forensic Analysis}}.
\newblock \bibinfo{journal}{The International Journal of Forensic Science} ,
  \bibinfo{pages}{35--44}\DOIprefix\doi{10.5769/J201301005}.
\bibitem[{Berry et~al.(2017)Berry, Homan and Eitzman}]{Berry2017}
\bibinfo{author}{Berry, A.}, \bibinfo{author}{Homan, J.},
  \bibinfo{author}{Eitzman, R.}, \bibinfo{year}{2017}.
\newblock \bibinfo{title}{{Threat Research WannaCry Malware Profile}}.
\newblock \URLprefix
  \url{https://www.fireeye.com/blog/threat-research/2017/05/wannacry-malware-profile.html}.
\bibitem[{Bose(2018)}]{Bose2018}
\bibinfo{author}{Bose, M.}, \bibinfo{year}{2018}.
\newblock \bibinfo{title}{{A Complete Comparison of VMware and VirtualBox}}.
\newblock \URLprefix
  \url{https://www.nakivo.com/blog/vmware-vs-virtual-box-comprehensive-comparison/}.
\bibitem[{Bradley(2016)}]{Bradley2016}
\bibinfo{author}{Bradley, S.}, \bibinfo{year}{2016}.
\newblock \bibinfo{title}{{Information Security Reading Room Ransomware}}.
\newblock \bibinfo{type}{Technical Report}. SANS Institute.
\newblock \URLprefix
  \url{https://www.sans.org/reading-room/whitepapers/awareness/paper/37317}.
\bibitem[{Carvey and Casey(2009)}]{Carvey2009}
\bibinfo{author}{Carvey, H.}, \bibinfo{author}{Casey, E.},
  \bibinfo{year}{2009}.
\newblock \bibinfo{title}{{Windows Forensic analysis DVD toolkit}}.
\newblock \bibinfo{edition}{2nd} ed., \bibinfo{publisher}{Burlington, MA :
  Syngress}.
\bibitem[{CERT-EU(2017)}]{CERT-EU2017}
\bibinfo{author}{CERT-EU}, \bibinfo{year}{2017}.
\newblock \bibinfo{title}{{WannaCry Ransomware Campaign Exploiting SMB
  Vulnerability}}.
\newblock \bibinfo{type}{Technical Report}. Computer Emergency Response Team -
  EU.
\newblock \URLprefix
  \url{https://cert.europa.eu/static/SecurityAdvisories/2017/CERT-EU-SA2017-012.pdf}.
\bibitem[{Comodo(2018)}]{Comodo2018}
\bibinfo{author}{Comodo}, \bibinfo{year}{2018}.
\newblock \bibinfo{title}{{8 Ransomware Attacks that have Occurrred Recently}}.
\newblock \URLprefix
  \url{https://enterprise.comodo.com/forensic-analysis/ransomware-attacks.php}.
\bibitem[{Dinaburg et~al.(2008)Dinaburg, Royal, Sharif and Lee}]{Dinaburg2008}
\bibinfo{author}{Dinaburg, A.}, \bibinfo{author}{Royal, P.},
  \bibinfo{author}{Sharif, M.}, \bibinfo{author}{Lee, W.},
  \bibinfo{year}{2008}.
\newblock \bibinfo{title}{{Ether: Malware Analysis via Hardware Virtualization
  Extension}}.
\newblock \bibinfo{journal}{Operating Systems]: Security and Protection} ,
  \bibinfo{pages}{51--62}\URLprefix
  \url{http://ether.gtisc.gatech.edu/ether{\_}ccs{\_}2008.pdf}.
\bibitem[{Egele et~al.(2012)Egele, Scholte, Kirda and Kruegel}]{Egele2012}
\bibinfo{author}{Egele, M.}, \bibinfo{author}{Scholte, T.},
  \bibinfo{author}{Kirda, E.}, \bibinfo{author}{Kruegel, C.},
  \bibinfo{year}{2012}.
\newblock \bibinfo{title}{{A survey on automated dynamic malware-analysis
  techniques and tools}}.
\newblock \bibinfo{journal}{ACM Computing Surveys} \bibinfo{volume}{44}.
\newblock \DOIprefix\doi{10.1145/2089125.2089126}.
\bibitem[{Europol(2016)}]{Europol2016}
\bibinfo{author}{Europol}, \bibinfo{year}{2016}.
\newblock \bibinfo{title}{{INTERNET ORGANISED CRIME 2016 IOCTA}}.
\newblock \bibinfo{type}{Technical Report}. Europol.
\newblock \URLprefix
  \url{https://www.europol.europa.eu/activities-services/main-reports/internet-organised-crime-threat-assessment-iocta-2016},
  \DOIprefix\doi{10.2813/275589}.
\bibitem[{Europol(2018)}]{Europol2018}
\bibinfo{author}{Europol}, \bibinfo{year}{2018}.
\newblock \bibinfo{title}{{INTERNET ORGANISED CRIME 2018 IOCTA}}.
\newblock \bibinfo{type}{Technical Report}. Europol.
\newblock \URLprefix
  \url{https://www.europol.europa.eu/internet-organised-crime-threat-assessment-2018},
  \DOIprefix\doi{10.2813/858843}.
\bibitem[{Halderman et~al.(2009)Halderman, Schoen, Heninger, Clarkson, Paul,
  Calandrino, Feldman, Appelbaum and Felten}]{Halderman2009}
\bibinfo{author}{Halderman, J.A.}, \bibinfo{author}{Schoen, S.D.},
  \bibinfo{author}{Heninger, N.}, \bibinfo{author}{Clarkson, W.},
  \bibinfo{author}{Paul, W.}, \bibinfo{author}{Calandrino, J.A.},
  \bibinfo{author}{Feldman, A.J.}, \bibinfo{author}{Appelbaum, J.},
  \bibinfo{author}{Felten, E.W.}, \bibinfo{year}{2009}.
\newblock \bibinfo{title}{{Lest We Remember : Cold Boot Attacks on Encryption
  Keys}}.
\newblock \bibinfo{journal}{Communications of the ACM,} \bibinfo{volume}{52},
  \bibinfo{pages}{91--98}.
\newblock \DOIprefix\doi{10.1145/1506409.1506429}.
\bibitem[{Hargreaves and Chivers(2008)}]{Hargreaves2008}
\bibinfo{author}{Hargreaves, C.}, \bibinfo{author}{Chivers, H.},
  \bibinfo{year}{2008}.
\newblock \bibinfo{title}{{Recovery of encryption keys from memory using a
  linear scan}}.
\newblock \bibinfo{journal}{ARES 2008 - 3rd International Conference on
  Availability, Security, and Reliability, Proceedings} ,
  \bibinfo{pages}{1369--1376}\DOIprefix\doi{10.1109/ARES.2008.109}.
\bibitem[{Hautala(2019)}]{Hautala2019}
\bibinfo{author}{Hautala, L.}, \bibinfo{year}{2019}.
\newblock \bibinfo{title}{{States brace for ransomware assaults on voter
  registries}}.
\newblock \URLprefix
  \url{https://www.cnet.com/news/wi-fi-6-is-barely-here-but-wi-fi-7-is-already-on-the-way/}.
\bibitem[{Heninger and Feldman(2008)}]{Heninger2008}
\bibinfo{author}{Heninger, N.}, \bibinfo{author}{Feldman, A.},
  \bibinfo{year}{2008}.
\newblock \bibinfo{title}{{AESKeyFind}}.
\newblock \URLprefix
  \url{https://github.com/eugenekolo/sec-tools/tree/master/crypto/aeskeyfind/aeskeyfind}.
\bibitem[{Hoopes(2009)}]{Hoopes2009}
\bibinfo{author}{Hoopes, J.}, \bibinfo{year}{2009}.
\newblock \bibinfo{title}{{Chapter 6. Malware Analysis Solutions}}, in:
  \bibinfo{booktitle}{Virtualization Security Protecting Virtualized
  Environments}. \bibinfo{edition}{1st} ed.. \bibinfo{publisher}{O'Reilly}.
  chapter \bibinfo{chapter}{Chapter 6.}
\newblock \URLprefix
  \url{https://learning.oreilly.com/library/view/virtualization-for-security/9781597493055/{\#}toc}.
\bibitem[{Intelligence and Analysis(2019)}]{Intelligence2019}
\bibinfo{author}{Intelligence, T.}, \bibinfo{author}{Analysis, I.},
  \bibinfo{year}{2019}.
\newblock \bibinfo{title}{{2019 SonicWall Cyber Threat Report}}.
\newblock \bibinfo{type}{Technical Report} \bibinfo{number}{July}. SonicWall.
\newblock \URLprefix \url{www.sonicwall.com}.
\bibitem[{Issa(2019)}]{Issa}
\bibinfo{author}{Issa, J.}, \bibinfo{year}{2019}.
\newblock \bibinfo{title}{{A deep dive into Phobos ransomware}}.
\newblock \URLprefix
  \url{https://blog.malwarebytes.com/threat-analysis/2019/07/a-deep-dive-into-phobos-ransomware/}.
\bibitem[{Kapersky(2018)}]{Kapersky2018}
\bibinfo{author}{Kapersky}, \bibinfo{year}{2018}.
\newblock \bibinfo{title}{{Top 5 most notorious cyberattacks}}.
\newblock \URLprefix
  \url{https://www.kaspersky.com/blog/five-most-notorious-cyberattacks/24506/}.
\bibitem[{Kornblum(2019)}]{Kornblum}
\bibinfo{author}{Kornblum, J.}, \bibinfo{year}{2019}.
\newblock \bibinfo{title}{findaes}.
\newblock \URLprefix \url{http://jessekornblum.com/tools/findaes/}.
\bibitem[{Ligh et~al.(2014)Ligh, Case, Levy and Walters}]{Ligh2014}
\bibinfo{author}{Ligh, M.H.}, \bibinfo{author}{Case, A.},
  \bibinfo{author}{Levy, J.}, \bibinfo{author}{Walters, A.},
  \bibinfo{year}{2014}.
\newblock \bibinfo{title}{{The Art of memory Forensics}}.
\newblock \bibinfo{publisher}{Wiley}.
\bibitem[{Maartmann-Moe et~al.(2009)Maartmann-Moe, Thorkildsen and
  {\AA}rnes}]{Maartmann-Moe2009}
\bibinfo{author}{Maartmann-Moe, C.}, \bibinfo{author}{Thorkildsen, S.E.},
  \bibinfo{author}{{\AA}rnes, A.}, \bibinfo{year}{2009}.
\newblock \bibinfo{title}{{The persistence of memory: Forensic identification
  and extraction of cryptographic keys}}.
\newblock \bibinfo{journal}{DFRWS 2009 Annual Conference} \bibinfo{volume}{6},
  \bibinfo{pages}{132--140}.
\newblock \DOIprefix\doi{10.1016/j.diin.2009.06.002}.
\bibitem[{Malwarebytes(2019)}]{Malwarebytes2019}
\bibinfo{author}{Malwarebytes}, \bibinfo{year}{2019}.
\newblock \bibinfo{title}{{Cybercrime Tactics and Techniques Q1 2019}}.
\newblock \bibinfo{type}{Technical Report}. MalwareBytes.
\newblock \URLprefix
  \url{https://www.malwarebytes.com/pdf/labs/Cybercrime-Tactics-and-Techniques-Q1-2017.pdf}.
\bibitem[{Mamedov et~al.(2018)Mamedov, Sinitsyn and Ivanov}]{Mamedov}
\bibinfo{author}{Mamedov, O.}, \bibinfo{author}{Sinitsyn, F.},
  \bibinfo{author}{Ivanov, A.}, \bibinfo{year}{2018}.
\newblock \bibinfo{title}{{Bad Rabbit ransomware}}.
\newblock \URLprefix \url{https://securelist.com/bad-rabbit-ransomware/82851/}.
\bibitem[{{McAfee Labs}(2016)}]{McAfeeLabs2016}
\bibinfo{author}{{McAfee Labs}}, \bibinfo{year}{2016}.
\newblock \bibinfo{title}{{Understanding Ransomware and Strategies to Defeat
  it}}.
\newblock \bibinfo{journal}{Network Security} ,
  \bibinfo{pages}{1--18}\URLprefix
  \url{https://www.mcafee.com/us/resources/white-papers/wp-understanding-ransomware-strategies-defeat.pdf}.
\bibitem[{McLaren et~al.(2019a)McLaren, Buchanan, Russell and
  Tan}]{McLaren2019a}
\bibinfo{author}{McLaren, P.}, \bibinfo{author}{Buchanan, W.J.},
  \bibinfo{author}{Russell, G.}, \bibinfo{author}{Tan, Z.},
  \bibinfo{year}{2019}a.
\newblock \bibinfo{title}{{Deriving ChaCha20 Key Streams From Targeted Memory
  Analysis}}.
\newblock \bibinfo{journal}{Journal of Information Security and Applications}
  \bibinfo{volume}{48}.
\newblock \URLprefix
  \url{http://arxiv.org/abs/1907.11941{\%}0Ahttp://dx.doi.org/10.1016/j.jisa.2019.102372},
  \DOIprefix\doi{10.1016/j.jisa.2019.102372},
  \href{http://arxiv.org/abs/1907.11941}{\tt arXiv:1907.11941}.
\bibitem[{McLaren et~al.(2019b)McLaren, Russell, Buchanan and
  Tan}]{McLaren2019}
\bibinfo{author}{McLaren, P.}, \bibinfo{author}{Russell, G.},
  \bibinfo{author}{Buchanan, W.J.}, \bibinfo{author}{Tan, Z.},
  \bibinfo{year}{2019}b.
\newblock \bibinfo{title}{{Decrypting live SSH traffic in virtual
  environments}}.
\newblock \bibinfo{journal}{Digital Investigation} \bibinfo{volume}{29},
  \bibinfo{pages}{109--117}.
\newblock \DOIprefix\doi{10.1016/j.diin.2019.03.010},
  \href{http://arxiv.org/abs/arXiv:1907.10835v1}{\tt arXiv:arXiv:1907.10835v1}.
\bibitem[{Mekynyk et~al.(2019)Mekynyk, Speier-Pero and Connors}]{Mekynyk2019}
\bibinfo{author}{Mekynyk, S.A.}, \bibinfo{author}{Speier-Pero, C.},
  \bibinfo{author}{Connors, E.}, \bibinfo{year}{2019}.
\newblock \bibinfo{title}{{Blockchain is Vastly Overrated; Supply Chain Cyber
  Security is Vastly Underrated}}.
\newblock \bibinfo{journal}{Supply Chain Management Review}
  \bibinfo{volume}{June}.
\newblock \URLprefix \url{scmr.com}.
\bibitem[{Nissim et~al.(2019)Nissim, Lahav, Cohen, Elovici and
  Rokach}]{Nissim2019}
\bibinfo{author}{Nissim, N.}, \bibinfo{author}{Lahav, O.},
  \bibinfo{author}{Cohen, A.}, \bibinfo{author}{Elovici, Y.},
  \bibinfo{author}{Rokach, L.}, \bibinfo{year}{2019}.
\newblock \bibinfo{title}{{Volatile memory analysis using the MinHash method
  for efficient and secured detection of malware in private cloud}}.
\newblock \bibinfo{journal}{Computers {\&} Security} \bibinfo{volume}{87}.
\bibitem[{O'Donnall(2019)}]{ODonnall2019}
\bibinfo{author}{O'Donnall, L.}, \bibinfo{year}{2019}.
\newblock \bibinfo{title}{{Coordinated Ransomware Attack Hits 23 Texas
  Government Agencies}}.
\newblock \URLprefix
  \url{https://threatpost.com/coordinated-ransomware-attack-hits-23-texas-government-agencies/147457/}.
\bibitem[{O'Donnell(2020)}]{ODonnell2020}
\bibinfo{author}{O'Donnell, L.}, \bibinfo{year}{2020}.
\newblock \bibinfo{title}{{ThreatList: Ransomware Costs Double in Q4,
  Sodinokibi Dominates}}.
\newblock \URLprefix
  \url{https://threatpost.com/threatlist-ransomware-costs-double-in-q4-sodinokibi-dominates/152200/}.
\bibitem[{{Panda Security}(2017)}]{Security2017}
\bibinfo{author}{{Panda Security}}, \bibinfo{year}{2017}.
\newblock \bibinfo{title}{{Technical Analysis of Bad Rabbit}}.
\newblock \bibinfo{journal}{Panda Security} \bibinfo{volume}{5},
  \bibinfo{pages}{1--5}.
\bibitem[{Perekalin(2018)}]{Perekalin2018}
\bibinfo{author}{Perekalin, A.}, \bibinfo{year}{2018}.
\newblock \bibinfo{title}{{Bad Rabbit: A new ransomware epidemic is on the
  rise}}.
\newblock \URLprefix
  \url{https://www.kaspersky.com/blog/bad-rabbit-ransomware/19887/}.
\bibitem[{Ptacek(2008)}]{Ptacek2008}
\bibinfo{author}{Ptacek, T.}, \bibinfo{year}{2008}.
\newblock \bibinfo{title}{{Recover a Private Key from Process Memory}}.
\newblock \URLprefix
  \url{http://www.matasano.com/log/178/recovera-{\%}0Aprivate-key-from-process-memory}.
\bibitem[{Rossow et~al.(2012)Rossow, Dietrich, Grier, Kreibich, Paxson,
  Pohlmann, Bos and {Van Steen}}]{Rossow2012}
\bibinfo{author}{Rossow, C.}, \bibinfo{author}{Dietrich, C.J.},
  \bibinfo{author}{Grier, C.}, \bibinfo{author}{Kreibich, C.},
  \bibinfo{author}{Paxson, V.}, \bibinfo{author}{Pohlmann, N.},
  \bibinfo{author}{Bos, H.}, \bibinfo{author}{{Van Steen}, M.},
  \bibinfo{year}{2012}.
\newblock \bibinfo{title}{{Prudent practices for designing malware experiments:
  Status quo and outlook}}.
\newblock \bibinfo{journal}{Proceedings - IEEE Symposium on Security and
  Privacy} , \bibinfo{pages}{65--79}\DOIprefix\doi{10.1109/SP.2012.14}.
\bibitem[{Ruff(2008)}]{Ruff2008}
\bibinfo{author}{Ruff, N.}, \bibinfo{year}{2008}.
\newblock \bibinfo{title}{{Windows memory forensics}}.
\newblock \bibinfo{journal}{Journal in Computer Virology} \bibinfo{volume}{4},
  \bibinfo{pages}{83--100}.
\newblock \DOIprefix\doi{10.1007/s11416-007-0070-0}.
\bibitem[{Salvi(2015)}]{Salvi2015}
\bibinfo{author}{Salvi, H.U.}, \bibinfo{year}{2015}.
\newblock \bibinfo{title}{{Ransomware : A Cyber Extortion}}.
\newblock \bibinfo{journal}{Asian Journal of Convergence in Technology}
  \bibinfo{volume}{II}.
\bibitem[{Sanabria(2007)}]{Sanabria2007}
\bibinfo{author}{Sanabria, A.}, \bibinfo{year}{2007}.
\newblock \bibinfo{title}{{Malware Analysis : Environment Design and
  Architecture}}.
\newblock \bibinfo{type}{Technical Report}. SANS Institute.
\newblock \URLprefix
  \url{https://www.sans.org/reading-room/whitepapers/threats/paper/1841}.
\bibitem[{Saravanan and Mukesh(2014)}]{Saravanan2014}
\bibinfo{author}{Saravanan, M.}, \bibinfo{author}{Mukesh, K.},
  \bibinfo{year}{2014}.
\newblock \bibinfo{title}{{Forensic Recovery of Fully Encrypted Volume}}.
\newblock \bibinfo{journal}{International Journal of Computer Applications}
  \bibinfo{volume}{91}, \bibinfo{pages}{18--21}.
\newblock \DOIprefix\doi{10.5120/15892-4896}.
\bibitem[{Shamir and {Van Someren}(1998)}]{Shamir1998}
\bibinfo{author}{Shamir, A.}, \bibinfo{author}{{Van Someren}, N.},
  \bibinfo{year}{1998}.
\newblock \bibinfo{title}{{Playing ‘hide and seek' with stored keys}}.
\newblock \bibinfo{journal}{Lecture Notes in Computer Science (including
  subseries Lecture Notes in Artificial Intelligence and Lecture Notes in
  Bioinformatics)} \bibinfo{volume}{1648}, \bibinfo{pages}{118--124}.
\bibitem[{Sikorski and Hong(2012)}]{Sikorski2012}
\bibinfo{author}{Sikorski, A.}, \bibinfo{author}{Hong, A.},
  \bibinfo{year}{2012}.
\newblock \bibinfo{title}{{Practical Malware Analysis}}.
\newblock \bibinfo{publisher}{No Starch Press}, \bibinfo{address}{San
  Francisco}.
\bibitem[{SonicWall(2019)}]{SonicWall2019}
\bibinfo{author}{SonicWall}, \bibinfo{year}{2019}.
\newblock \bibinfo{title}{{Unmasking the threats that target glkobal
  enterprises, governments abd SMBs}}.
\newblock \bibinfo{journal}{Journal of Chemical Information and Modeling}
  \bibinfo{volume}{53}.
\bibitem[{Sood and Hurley(2017)}]{Sood2017}
\bibinfo{author}{Sood, K.}, \bibinfo{author}{Hurley, S.}, \bibinfo{year}{2017}.
\newblock \bibinfo{title}{{NotPetya Technical Analysis – A Triple Threat:
  File Encryption, MFT Encryption, Credential Theft}}.
\newblock \URLprefix
  \url{https://www.crowdstrike.com/blog/petrwrap-ransomware-technical-analysis-triple\\
  -threat-file-encryption-mft-encryption-credential-theft/}.
\bibitem[{Sophos(2019)}]{Sophos2019}
\bibinfo{author}{Sophos}, \bibinfo{year}{2019}.
\newblock \bibinfo{title}{{Ransomware: How an attack works - Sophos
  Community}}.
\newblock \bibinfo{journal}{Sophos} , \bibinfo{pages}{1--2}\URLprefix
  \url{https://community.sophos.com/kb/en-us/124699}.
\bibitem[{Trenholme()}]{Trenholme}
\bibinfo{author}{Trenholme, S.}, .
\newblock \bibinfo{title}{{Rijndael's key schedule}}.
\newblock \URLprefix \url{https://www.samiam.org/key-schedule.html}.
\bibitem[{Trenholme(2014)}]{Trenholme2014}
\bibinfo{author}{Trenholme, S.}, \bibinfo{year}{2014}.
\newblock \bibinfo{title}{findaes}.
\newblock \URLprefix \url{https://sourceforge.net/projects/findaes/}.
\bibitem[{Vanderburg(2019)}]{Vanderburg2019}
\bibinfo{author}{Vanderburg, E.}, \bibinfo{year}{2019}.
\newblock \bibinfo{title}{{A Timeline of Ransomware Advances}}.
\newblock \URLprefix \url{https://www.tcdi.com/ransomware-timeline/}.
\bibitem[{{Vipre Security}(2017)}]{VipreSecurity2017}
\bibinfo{author}{{Vipre Security}}, \bibinfo{year}{2017}.
\newblock \bibinfo{title}{{WannaCry Technical Analysis : Support}}.
\newblock \URLprefix
  \url{https://support.threattracksecurity.com/support/solutions/articles/1000250396-wannacry-technical-analysis}.
\bibitem[{Volatility(2019)}]{Volatility2019}
\bibinfo{author}{Volatility}, \bibinfo{year}{2019}.
\newblock \bibinfo{title}{{Volatility Foundation}}.
\newblock \URLprefix \url{https://www.volatilityfoundation.org/}.
\bibitem[{Walters and Petroni(2007)}]{Walters2007}
\bibinfo{author}{Walters, A.}, \bibinfo{author}{Petroni, N.L.},
  \bibinfo{year}{2007}.
\newblock \bibinfo{title}{{Volatools: Integrating Volatile Memory Forensics
  into the Digital Investigation Process}}.
\newblock \bibinfo{journal}{Black Hat DC} , \bibinfo{pages}{1--18}.

\end{thebibliography}

\end{document}